\documentclass[twocolumn,showpacs,preprintnumbers,amsmath,amssymb,prb]{revtex4}

\usepackage{graphicx}
\usepackage{color}

\begin{document}

\title{Thermal properties of graphene from path-integral simulations}
\author{Carlos P. Herrero}
\author{Rafael Ram\'irez}
\affiliation{Instituto de Ciencia de Materiales de Madrid,
         Consejo Superior de Investigaciones Cient\'ificas (CSIC),
         Campus de Cantoblanco, 28049 Madrid, Spain }
\date{\today}

\begin{abstract}
Thermal properties of graphene monolayers are studied by path-integral 
molecular dynamics (PIMD) simulations, which take into account the 
quantization of vibrational modes in the crystalline membrane, and allow 
one to consider anharmonic effects in these properties. 
This system was studied at temperatures in the range from 12 to 2000~K 
and zero external stress, by describing the interatomic interactions 
through the LCBOPII effective potential.
We analyze the internal energy and specific heat and compare the results
derived from the simulations with those yielded by a harmonic approximation
for the vibrational modes. This approximation turns out to be rather 
precise up to temperatures of about 400~K.
At higher temperatures, we observe an influence of the elastic energy,
due to the thermal expansion of the graphene sheet.
Zero-point and thermal effects on the in-plane and ``real'' surface of 
graphene are discussed. The thermal expansion coefficient $\alpha$ of 
the real area is found to be positive at all temperatures, 
in contrast to the expansion coefficient $\alpha_p$ of the 
in-plane area, which is negative at low temperatures, and becomes positive
for $T \gtrsim$ 1000~K. 
\end{abstract}



\maketitle

\section{Introduction}

  Graphene has been extensively studied in last years, not only for its 
remarkable electronic properties,\cite{ge07,ca09b,fl11} but also for other
features such as elastic and thermal properties.
In fact, graphene displays high values of thermal 
conductivity,\cite{gh08b,ni09,ba11} as well as large in-plane elastic 
constants.\cite{le08}
The optimum structural arrangement for pure defect-free graphene
corresponds to a honeycomb lattice, but
departures from this flat structure may significantly affect
both its atomic-scale and macroscopic properties.\cite{me07}

There are several reasons for a graphene sheet to bend and depart from 
absolute planarity, such as the presence of defects and
external stresses.\cite{fa07,an12}
For a perfect two-dimensional (2D) crystalline layer in 
three-dimensional (3D) space, thermal fluctuations at finite temperatures 
cause out-of-plane motion of the carbon atoms,
and even for $T \to 0$, quantum fluctuations associated to zero-point 
motion yield a departure of strict planarity of the graphene sheet.\cite{he16}

Understanding structural and thermal properties of 2D systems in 3D space
is a challenge in modern statistical physics.\cite{sa94,ne04,ta13}
This question has been mainly treated in connection with biological
membranes and soft condensed matter.\cite{ta13,ch15,ru12} 
However, the large complexity of these systems has limited the development
of microscopic approaches based on realistic interatomic interactions.
Graphene, as a well-characterized crystalline membrane, 
can be considered as a model system where an atomistic description is 
possible, paving the path to a better understanding of the physical 
properties of this kind of systems.
In this line, the interest on thermal properties of graphene
has risen in the last few years,\cite{am14,po12b,wa16,fo13,ma16}
as is the case of thermal expansion and heat conduction,
which have been recently studied both experimentally and 
theoretically.\cite{ba11,po12b,al13,al14,si14,bu16}

Monte Carlo and molecular dynamics simulations have been employed
to study finite-temperature properties of graphene.
These simulations were based on 
{\em ab-initio},\cite{an12b,sh08,ch14}
tight-binding,\cite{ak12,ca09c,le13,he09a}
and empirical interatomic potentials.\cite{fa07,sh13,ra16,ma14,lo16}
In most cases, carbon atoms were described as classical particles,
which is reliable at relatively high temperatures (in the order of the
Debye temperature of the material), but is not suitable to study
thermodynamic variables at low temperature.
To take into account the quantum nature of the atomic motion,
path-integral simulations are well-suited, since in this procedure
nuclear degrees of freedom may be quantized, allowing one to include 
quantum and thermal fluctuations in many-body systems
at finite temperatures.\cite{gi88,ce95}  
Thus, path-integral simulations of a single graphene layer have been
recently carried out to study equilibrium properties 
of this material.\cite{br15,he16}
In addition to this, nuclear quantum effects have been studied earlier 
by using a combination of density-functional theory and a quasi-harmonic 
approximation for vibrational modes in this crystalline
membrane.\cite{mo05,sh12}

In this paper, the path-integral molecular dynamics (PIMD) method is
used to study thermal properties of graphene at temperatures
between 12 and 2000~K.
Simulation cells of different sizes are considered, as finite-size effects 
have been found earlier to be important for some equilibrium
properties of graphene.\cite{ga14,lo16,he16}
We study the thermal behavior of the graphene surface, taking into
account the difference between real surface and projected in-plane area. 
Particular emphasis is laid on the temperature dependence of the specific
heat at low $T$, for which results of the simulations are compared with
predictions based on harmonic vibrations of the crystalline membrane.

 The paper is organized as follows. In Sec.\,II, we describe the
computational method employed in the simulations.
Results for the internal energy of graphene are given in Sec.~III.
Structural properties such as the in-plane and real area are
discussed in Sec.~IV. In Sec.~V we present results for the specific
heat, and in Sec.\,VI we summarize the main results.

\section{Computational Method}

In this paper we use PIMD simulations to study equilibrium properties of 
graphene monolayers as a function of temperature.
The PIMD method is based on the path-integral formulation of statistical
mechanics,\cite{fe72} which has turned out to be a convenient 
nonperturbative approach 
to study finite-temperature properties of many-body quantum systems.
In this method, the partition function is evaluated  through
a discretization of the density matrix along cyclic paths, formed by 
a finite number $N_{\rm Tr}$ (Trotter number) of steps.\cite{fe72,kl90}
In actual applications of this procedure to numerical simulations,
this discretization causes the appearance of
$N_{\rm Tr}$ replicas (or beads) for each quantum particle. 
These replicas are treated in the calculations as classical particles, 
since the partition function of the real quantum system is isomorph to
that of a classical one, obtained by substituting each quantum particle 
by a ring polymer composed of $N_{\rm Tr}$ particles.\cite{gi88,ce95}
The dynamics in this procedure is artificial, as it
does not reflect the real quantum dynamics of the actual particles.
Nevertheless, it is useful for effectively sampling the
many-body configuration space, yielding accurate results for 
time-independent equilibrium properties of the quantum system under
consideration.
Details on this simulation technique can be found
elsewhere.\cite{ch81,gi88,ce95,he14}

An important point for this kind of simulations is a reliable description 
of the interatomic interactions, which have to be as realistic as possible.
We obtain a Born-Oppenheimer surface for the nuclear dynamics from 
the LCBOPII effective potential,\cite{lo05} as using an 
{\em ab-initio} method would largely reduce the size of the simulation 
cells to be handled. 
This is a long-range carbon bond order potential, which has been used  
to perform classical simulations of carbon-based systems, such
as diamond,\cite{lo05} graphite,\cite{lo05}, liquid carbon,\cite{gh05} 
and more recently graphene.\cite{fa07,za09,lo16}
It has been used, in particular, to study the carbon phase diagram including 
graphite, diamond, and the liquid, and displayed its accuracy in a comparison 
of the predicted graphite-diamond line with experimental results.\cite{gh05b}
The LCBOPII potential has been also found to describe well several 
properties of graphene, such as Young's modulus.\cite{za09,po12,ra17} 
In line with earlier simulations,\cite{ra16,he16,ra17} the original LCBOPII 
parameterization was slightly modified to increase the zero-temperature 
bending constant from 1.1 eV to a more realistic value of 1.49 eV.\cite{la14}
This effective potential has been recently employed to carry out PIMD
simulations, which allowed to quantify the magnitude of quantum effects
in graphene monolayers by comparing with results of classical
simulations.\cite{he16}

The calculations presented here have been carried out in the 
isothermal-isobaric ensemble, where one fixes the number of carbon 
atoms ($N$), the applied stress (here $P = 0$), and the temperature ($T$).
The stress $P$, with units of force per unit length, coincides with the
so-called mechanical or frame tension $\tau$ in several 
papers.\cite{ra17,fo08,sh16}
For comparison, some PIMD simulations were also carried out with constant
projected area $A_p$ in the reference $(x, y)$ plane.
This ensemble is similar to the $NVT$ ensemble employed for 
simulations of 3D materials, $V$ being the volume.
We used effective algorithms for performing PIMD simulations, as those 
described in the literature.\cite{tu92,tu98,ma99,tu02}   In particular,
we employed staging variables\cite{tu93} to define the bead coordinates, and
the constant-temperature ensemble was generated by coupling chains
of four Nos\'e-Hoover thermostats.\cite{no84,ho85} 
For the isothermal-isobaric simulations,
an additional chain of four barostats was coupled to the area $A_p$
of the simulation box to give the required constant stress 
(here $P = 0$).\cite{tu98,he14}
The equations of motion were integrated by using the reversible reference 
system propagator algorithm (RESPA), which allows one to deal with 
different time steps for the integration of the fast and slow
degrees of freedom.\cite{ma96}
The time step $\Delta t$ associated to the interatomic forces was taken in 
the range between 0.5 and 1 fs, which turned out to be adequate for the 
interatomic interactions, atomic masses, and temperatures
considered here.
More details on this kind of PIMD simulations are given 
elsewhere.\cite{tu98,he06,he11}

We consider rectangular simulation cells with similar side lengths
$L_x$ and $L_y$ in the $x$ and $y$ directions of the $(x, y)$ reference 
plane, and periodic boundary conditions were assumed.
Sampling of the configuration space has been carried out at temperatures
between 12~K and 2000~K.
The Trotter number $N_{\rm Tr}$ was taken proportional to the inverse
temperature ($N_{\rm Tr} \propto 1/T$), so that $N_{\rm Tr} \, T$ = 6000~K,
which turns out to roughly keep a constant precision in the PIMD results
at different temperatures.\cite{he06,he11,ra12}
Cells of size up to 33600 atoms were considered for simulations at 
$T \geq$ 300~K, and at lower temperatures, smaller cells were considered
due to the fast increase in the number of beads $N_{\rm Tr}$ for 
the carbon atoms. Given a temperature, a typical simulation run consisted 
of $3 \times 10^5$ PIMD steps for system equilibration, followed by
$6 \times 10^6$ steps for the calculation of ensemble average properties.

\section{Internal energy}

At $T = 0$ we find with the LCBOPII potential in a classical approach
a strictly planar graphene surface with an interatomic distance of 
1.4199 \AA, i.e., an area of 2.6189 \AA$^2$ per atom, which we call $A_0$.  
This corresponds to a graphene sheet with fixed atomic nuclei on their
equilibrium sites without spatial delocalization, giving the minimum 
energy $E_0$, taken as a reference for our calculations at
finite temperatures.
In a more realistic quantum approach, the low-temperature limit includes 
out-of-plane atomic fluctuations associated to zero-point motion, 
and the graphene layer is not strictly planar. 
In addition to this, anharmonicity of in-plane vibrations causes a small 
zero-point lattice expansion, yielding an interatomic distance
of 1.4287 \AA, i.e., about 1\% larger than the classical minimum.

At $P$ = 0, the internal energy $E$ is obtained as a sum of the kinetic and
potential energy obtained from the simulations at a given temperature.
The kinetic energy was calculated by using the virial
estimator,\cite{he82,tu98} which is known to have a statistical uncertainty
smaller than the potential energy of the system.

Since we are interested in the large-size (thermodynamic) limit 
of the variables considered here, it is important to reduce as much as
possible the finite-size effects associated to them.  Thus, we have 
corrected for the center-of-mass translational energy, a classical magnitude 
amounting to $E_{\rm CM} = 3 k_B T / 2$ at temperature $T$, and
that is usually neglected as an unimportant quantity in this context.
When considering the energy per atom, this quantity becomes irrelevant
for large systems, but in general one has to include it to accelerate
the convergence of the internal energy per atom with system size.
Then, we have added $3 k_B T / 2 N$ to the internal energy per atom obtained 
in PIMD simulations.

\begin{figure}
\vspace{-1.0cm}
\includegraphics[width=8.5cm]{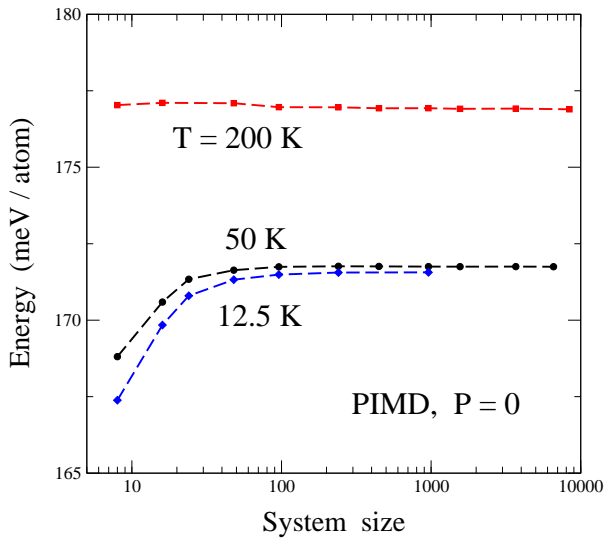}
\vspace{-0.5cm}
\caption{Internal energy per atom vs system size, as derived from PIMD
simulations at $P$ = 0. From top to bottom: $T$ = 200, 50, and 12.5 K.
Dashed lines are guides to the eye.
Error bars are less than the symbol size.
}
\label{f1}
\end{figure}

In our simulations of graphene, both kinetic and potential energy were
found to slightly increase with system size, but their convergence is
rather fast.
In Fig.~1 we present the internal energy per atom derived from 
PIMD simulations as a function of cell size at three temperatures:
200 K (squares), 50 K (circles), and 12.5 K (diamonds).
At the lowest temperature,
there appears a shift of about 4 meV/atom when increasing the cell
size from 8 atoms to the largest sizes considered here.
For cells in the order of 200 atoms the size effect in the internal energy
is almost inappreciable when compared to the largest cells.
The potential energy was found earlier to be smaller than the kinetic 
energy, indicating a nonnegligible anharmonicity of the
lattice vibrations.\cite{he16}
The convergence with system size becomes faster as the temperature is 
raised.    This is basically due to the fact that increasing 
the cell size effectively causes the appearance of low vibrational 
frequencies in the system, that
do not appear for smaller sizes. Increasing the temperature makes that
these new low-frequency modes behave ``more classically'' 
(see Sec.~V.A below).

\begin{figure}
\vspace{-1.0cm}
\includegraphics[width=8.5cm]{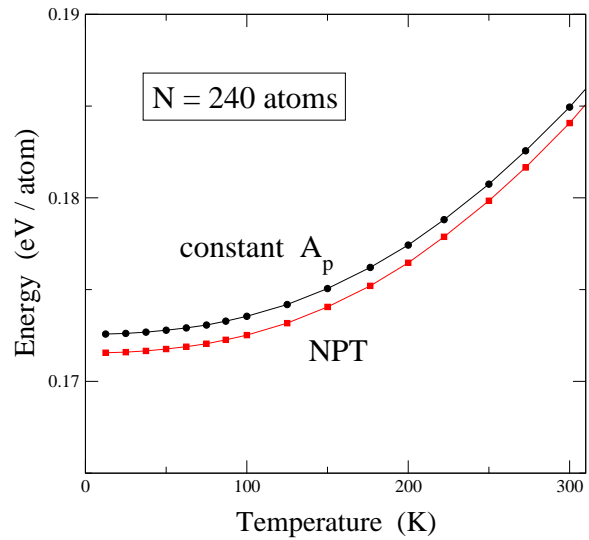}
\vspace{-0.5cm}
\caption{Internal energy obtained from PIMD simulations in the $NPT$ ensemble
($P = 0$) (circles) and in the constant-$A_p$ ensemble with $A_p = A_0$
(squares).
Note the zero-point energy, which amounts to about 0.17 eV/atom.
Error bars of the simulation data are smaller than the symbol size.
}
\label{f2}
\end{figure}

In Fig.~2 we display the temperature dependence of the internal energy,
as derived from PIMD in the $NPT$ isothermal-isobaric ensemble,
for system size $N$ = 240.
For comparison, we also present results obtained from constant-$A_p$ 
simulations with fixed area $A_p = A_0$.  The zero-point energy, 
$E_{\rm ZP}$, is found to be close to 0.17 eV/atom in both cases.
In the isobaric simulations, however, $E_{\rm ZP}$ is somewhat lower than
in the fixed-$A_p$ simulations, mainly due to the zero-point
expansion of the graphene layer with respect to the classical minimum
$A_0$. This expansion relaxes the compressive stress appearing for area 
$A_0$ in the presence of atomic quantum motion, and consequently the energy
decreases. The difference between zero-point energy in both cases amounts
to 1 meV/atom.  This difference between both sets of results decreases for 
rising temperature, and eventually the constant-area energy becomes lower 
than the $P = 0$ result for $T \gtrsim$ 1000 K (not shown in the figure).

\begin{figure}
\vspace{-1.0cm}
\includegraphics[width=8.5cm]{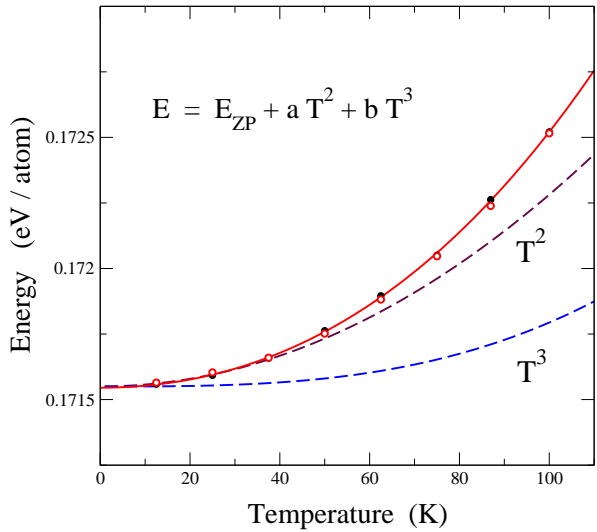}
\vspace{-0.5cm}
\caption{Temperature dependence of the internal energy of graphene
in the region up to 100 K. Symbols represent simulation results for
$P = 0$ obtained for $N = 240$ (solid circles) and $N = 960$
(open circles).
The line is a fit of the data points to the expression
$E = E_{\rm ZP} + a T^2 + b T^3$, with $E_{\rm ZP}$ the zero-point energy.
The dashed lines represent the contributions of the $T^2$ and $T^3$
terms to the fitted curve.
}
\label{f3}
\end{figure}

To better appreciate the low-temperature region, we present
in Fig.~3 the temperature dependence of the internal energy
obtained from PIMD simulations in the $NPT$ ensemble up to $T \sim$ 100 K. 
Symbols indicate results of the simulations for $N = 240$ (solid circles)
and $N = 960$ (open circles). Note that several solid circles are 
nearly unobservable, as they lie under the results for 960 atoms.
The solid line is a fit to the expression 
$E = E_{\rm ZP} + a T^2 + b T^3$, which displays a good agreement with
a temperature dependence of the energy in the region shown in Fig.~3. 
 For the coefficients $a$ and $b$ we found: 
 $a = 7.1 \times 10^{-8}$ eV K$^{-2}$
 and $b = 2.7 \times 10^{-10}$ eV K$^{-3}$.
Note that a linear term in this expression
for the internal energy is not possible for thermodynamic consistency,
since the specific heat $c_p = (\partial E / \partial T)_P$ has to vanish 
for $T \to 0$.  In Fig.~3 we also present separately the contributions 
of the $T^2$ and $T^3$ terms (dashed lines). 
The $T^2$ term is the main contribution to the energy
in the considered region, and controls the temperature dependence of
the energy up to about 40 K. This is important for the low-temperature
specific heat and will be further discussed in Sec.~V.

\section{Structural properties}

In our simulations in the isothermal-isobaric ensemble one fixes
the applied stress in the $(x, y)$ plane (here $P = 0$), 
allowing changes in the in-plane area of the simulation
cell for which periodic boundary conditions are applied.
Carbon atoms are free to move in the out-of-plane direction
($z$ coordinate), and in general any measure of the ``real'' surface 
of a graphene sheet at $T > 0$ should give a value larger than
the area of the simulation cell in the $(x, y)$ plane.
In this line, it has been argued for biological membranes that
their properties should be described using the concept 
of a real surface instead of a ``projected'' (in-plane)
surface.\cite{im06,wa09,ch15}  A similar question can be risen 
for crystalline membranes such as graphene.
This may be relevant for calculating thermodynamic properties, since
the in-plane area $A_p$ is the variable conjugate to the stress $P$
used in our simulations, and the real area (also called true, effective, 
or actual area in the literature\cite{im06,fo08,wa09,ch15})
is conjugate to the usually-called surface tension.\cite{sa94}
It is, in fact, the in-plane area $A_p$ which has been commonly employed 
in the literature to describe the results of atomistic simulations
of graphene layers.\cite{ga14,br15,za09,lo16,ch15} 
In the framework of biological membranes,
it was shown that values of the compressibility may
be very different when they are related to $A$ or to $A_p$, and
something similar has been recently found for the elastic properties of
graphene.\cite{ra17}

\begin{figure}
\vspace{-1.0cm}
\includegraphics[width=8.5cm]{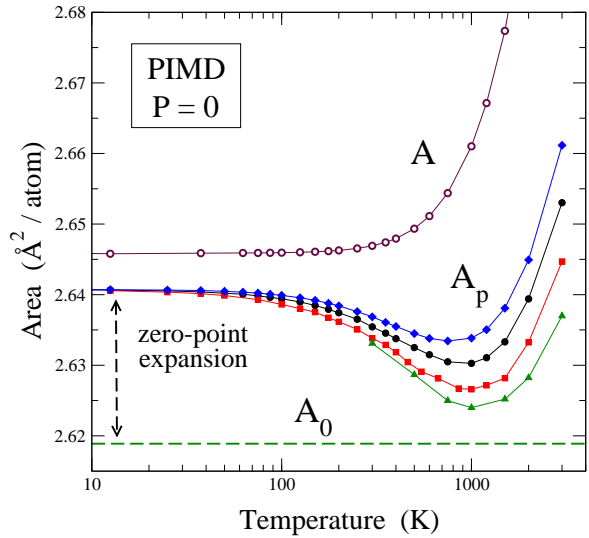}
\vspace{-0.5cm}
\caption{In-plane and real area vs temperature. Solid symbols represent
results for the in-plane area $A_p$, as derived from isothermal-isobaric
PIMD simulations for several system sizes. From top to bottom,
$N$ = 96 (diamonds), 240 (circles), 960 (squares), and 33600 (triangles).
Open circles correspond to the real area $A$ obtained for $N$ = 240 atoms;
other system sizes yielded results for $A$ indistinguishable from those
presented here.
Solid lines are guides to the eye.
Error bars are less than the symbol size.
The dashed line indicates the minimum-energy area $A_0$.
}
\label{f4}
\end{figure}

Here we calculate a real area $A$ in 3D space 
by a triangulation based on the actual atomic positions.
Each structural hexagon contributes to the area $A$ by a sum of six 
triangles, each one formed by the positions of two adjacent 
C atoms and the centroid (barycenter) of the hexagon
(mean position of the corresponding six vertices).\cite{ra17}
Other, qualitatively similar, definitions can be used for the real
area, such as those based on the interatomic distance C--C.\cite{ha16,he16}
In Fig.~4 we show the temperature dependence of the in-plane area 
$A_p$ and the real area $A$ of graphene, as derived from PIMD simulations,
in a semilogarithmic plot.
For $A_p$ we present results for various cell sizes as solid symbols. 
From top to bottom: $N$ = 96 (diamonds), 240 (circles), 960 (squares), 
and 33600 (triangles). 
Open circles represent results for the area $A$ obtained with 
$N$ = 240 atoms; results for larger cells are indistinguishable from them.
In fact, $A$ shows, in contrast to $A_p$, a small finite-size effect not
visible at the scale of Fig.~4. 
The horizontal dashed line in Fig.~4 indicates the minimum-energy area
$A_0$, corresponding to a planar classical sheet at $T = 0$.

For the area $A$ one observes a nearly constant value up to about 200 K,
followed by an increase at higher temperatures, similar to that observed
for the volume of 3D crystalline solids such as diamond.\cite{he00c}
The in-plane area $A_p$ decreases in the range from $T = 0$ to temperatures 
in the order of 1000 K, where it reaches a minimum, 
and then it increases at higher $T$.
Here, the finite-size effect is important in both the temperature
$T_m$ of the minimum and the value of the minimum area. 
For rising system size, the temperature $T_m$ shifts to higher values,
whereas the minimum $A_p$ decreases with increasing $N$.
These results for the in-plane area are reminiscent of those found 
from classical Monte Carlo and molecular dynamics simulations 
of graphene,\cite{za09,ga14,br15} but in PIMD simulations we find 
a more pronounced  decrease in $A_p$ in the temperature region 
from 0 to 1000 K. 

In the limit $T \to 0$, the areas $A$ and $A_p$ converge to 
2.6459 \AA$^2$ and 2.6407 \AA$^2$, respectively.
It is important to note that, in spite of the appreciable differences 
in the in-plane area per atom for the different system sizes,
all of them converge at low $T$ to the same value.
In the low-temperature region one observes first a zero-point expansion
of about 0.02 \AA$^2$ ($\sim$ 1\%), mainly due to an increase in the 
mean C--C bond length, caused by zero-point vibrations (an anharmonic
effect).  The small difference of a 0.2\% between real and in-plane areas 
is associated to out-of-plane zero-point motion, which causes that even at
$T = 0$ the layer is not strictly planar. Note that this is a pure 
quantum effect, since in classical simulations at $T \to 0$ 
one finds a planar layer in which $A$ and $A_p$ coincide.\cite{he16,ra17}

\begin{figure}
\vspace{-1.0cm}
\includegraphics[width=8.5cm]{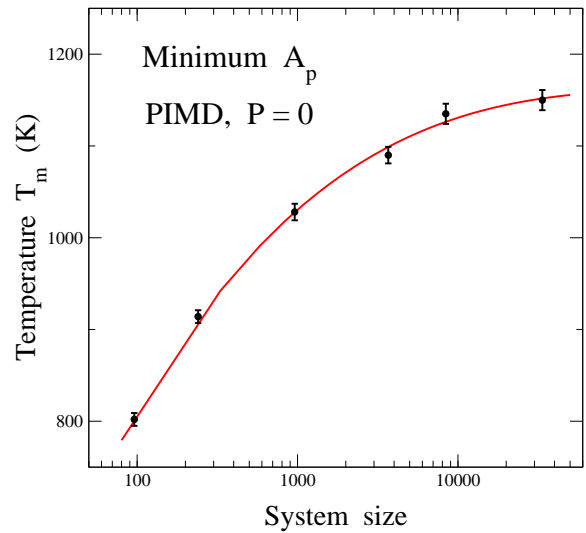}
\vspace{-0.5cm}
\caption{Temperature $T_m$ corresponding to the minimum in-plane area
$A_p$ vs system size $N$, as derived from PIMD simulations for $P = 0$.
The line is a polynomial fit the the data points.
}
\label{f5}
\end{figure}

Moreover, the temperature $T_m$ at which $A_p$ reaches its minimum is
size-dependent (see Fig.~4). In Fig.~5 we present the dependence of
$T_m$ on system size. Solid symbols indicate results of PIMD, and the
solid line is a polynomial fit to the data points.
It has been shown earlier from extensive classical simulations that
the in-plane area $A_p$ has an important finite-size effect, but its
large-$N$ limit is well defined.\cite{ra16,lo16}
Something similar is expected for the results of PIMD simulations, and
in particular for the temperature $T_m$.
Thus, in the large-$N$ limit, the size-dependent $T_m$ converges to 
a value $\lesssim$ 1200 K. A more precise result for this limit would 
require consideration of larger system sizes, not accessible at present
with our simulation procedure.

Beginning from $T = 0$,
the surface $A$ is larger than $A_p$, and the difference between both
increases with temperature. In fact, $A_p$ is the projection of $A$
on the $(x, y)$ plane, and the actual surface becomes increasingly 
bent as temperature is raised and out-of-plane atomic displacements 
are larger.  For the area $A$ we do not observe the decrease displayed 
by $A_p$.  Moreover, the areas $A$ and $A_p$ derived from PIMD 
simulations show a temperature derivative which approaches zero 
as $T \to 0$, in agreement with the third law of thermodynamics.

The behavior of $A_p$ as a function of $T$ is basically due to
a competition between two opposite factors.
First, the real area $A$ rises as $T$ is increased in the whole 
temperature range considered here.
Second, bending of the surface gives rise to a decrease in its 2D 
projection, $A_p$.
At $T \lesssim$ 1000~K, the decrease due to out-of-plane vibrations 
dominates the thermal expansion of the real surface, so that
$d A_p / d T < 0$.
At $T \gtrsim$ 1000~K, the increase in $A$ dominates the contraction
in the projected area associated to out-of-plane atomic displacements.
This behavior is qualitatively similar to that found from classical
molecular dynamics and Monte Carlo simulations, as well as analytical
calculations, where a minimum in the temperature dependence of $A_p$ 
was also found.\cite{za09,ga14,mi15b,he16}
The main difference is that the contraction of  $A_p$ respect the 
zero-temperature value is in the quantum case significantly larger 
than for classical calculations.

\begin{figure}
\vspace{-1.0cm}
\includegraphics[width=8.5cm]{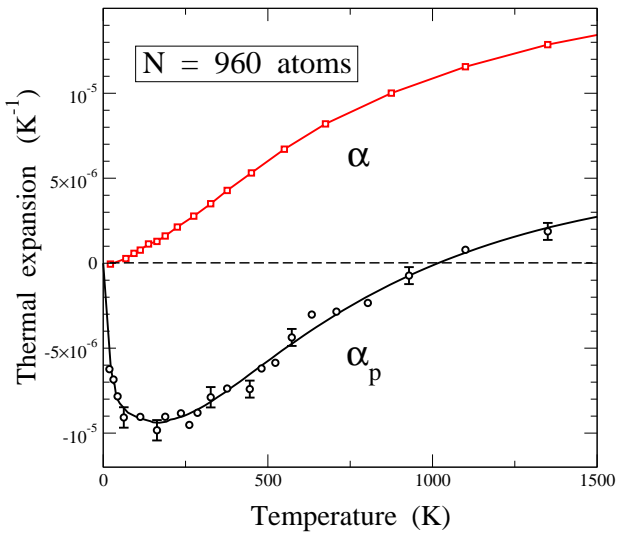}
\vspace{-0.5cm}
\caption{Thermal expansion coefficients $\alpha$ and $\alpha_p$ of
graphene vs temperature, as derived from PIMD simulations for a cell
including 960 atoms. Symbols are data points obtained from numerical
derivatives of $A$ and $A_p$ (squares, $\alpha$; circles, $\alpha_p$).
Solid lines indicate the temperature derivatives of polynomial
fits to the obtained values for $A$ and $A_p$.
Error bars for $\alpha$ are less than the symbol size.
}
\label{f6}
\end{figure}

According to our definitions of the areas $A$ and $A_p$,
we consider two different thermal expansion coefficients:
\begin{equation}
   \alpha = \frac{1}{A}  \left( \frac{\partial A}{\partial T} \right)_P
\end{equation}
and
\begin{equation}
 \alpha_p = \frac{1}{A_p}
         \left( \frac{\partial A_p}{\partial T} \right)_P   \, .
\end{equation}
The area $A$ derived from our PIMD simulations shows a negligible
size effect for $N \gtrsim$ 100 atoms, as indicated above. 
Hence, the same occurs
for the coefficient $\alpha$, which vanishes in the zero-temperature 
limit and turns out to be positive at all finite temperatures 
considered here.
In Fig.~6 we display the thermal expansion coefficients
$\alpha$ and $\alpha_p$ derived from our simulations.
Symbols are data points obtained from a numerical derivative of
the areas $A$ (squares) and $A_p$ (circles) corresponding to $N$ = 960.
For these derivatives we took temperature intervals ranging from 10 K at
low temperature to about 100 K at temperatures higher than 1000 K.
To check the precision of these numerical derivatives, we fitted the 
obtained values for $A$ and $A_p$ to polynomial expressions in the 
temperature range from $T$ = 10 to 1500 K. 
We then obtained the temperature derivative of these polynomials,
which yielded the solid lines displayed in Fig.~6. The agreement between
both procedures for calculating $\alpha$ and $\alpha_p$ is good.
Note that the noise in the values of $\alpha_p$ obtained from numerical
differences is clearly larger than that found for $\alpha$, as a consequence 
of the larger fluctuations in $A_p$.

Both $\alpha$ and $\alpha_p$ converge to zero in the low-temperature
limit. The behavior of $\alpha$ is similar to that observed for
crystalline materials, as indicated above for the temperature
dependence of the area $A$. 
However, $\alpha_p$ decreases fast for increasing temperature, until
reaching a minimum, which for $N = 960$ amounts to 
$\approx -9 \times 10^{-6}$ K$^{-1}$ at $T_m \sim$ 200 K. 
At higher $T$, $\alpha_p$ rises and becomes
positive at a temperature of about 1000 K (where $A_p$ reaches its
minimum value).
In view of the results for $A_p$ displayed in Fig.~4, 
the minimum value of $\alpha_p$ is expected to be size-dependent. 
Analyzing the results for the different system sizes studied here,
we estimate in the large-size limit a minimum value 
$\alpha_p = -1.1(\pm 0.1) \times 10^{-5}$ K$^{-1}$.
It is interesting to note that the difference $\alpha - \alpha_p$,
which vanishes at $T = 0$, increases fast as temperature is raised,
and takes a value  $\approx 10^{-5}$ K$^{-1}$ for 
temperatures higher than 1000 K.

Our results for $\alpha_p$ are qualitatively similar to those derived 
earlier from other theoretical techniques. 
Jiang {\em et al.}\cite{ji09} employed a 
nonequilibrium Green's function approach, and found for free-standing 
graphene a minimum $\approx 10^{-5}$ K$^{-1}$, close to our data shown 
in Fig.~6.  They found a crossover from negative to positive 
$\alpha_p$ at a temperature $T \approx$ 600 K, 
lower than our PIMD results.
Such a crossover at a point with $\alpha_p = 0$ was obtained by
da Silva {\em et al.}\cite{si14} at $T \approx$ 400~K from an 
unsymmetrized self-consistent-field method.
Experimental results at room temperature are not far from our data
at 300~K: $\alpha_p = - 8 \times  10^{-6}$ K$^{-1}$ derived from
Raman spectroscopy results,\cite{yo11} and $- 7 \times  10^{-6}$ K$^{-1}$
found from scanning electron microscopy.\cite{ba09}
The consistency between different measurements is not so good for the
temperature dependence of $\alpha_p$, and in fact the increase in
$\alpha_p$ at $T >$ 300~K is much faster in the former case \cite{yo11} 
than it the latter.\cite{ba09}

To make connection of our results derived from atomistic simulations 
with analytical formulations of membranes,
we note that the relation between $A$ and $A_p$ can be expressed
in the continuum limit (macroscopic view) as\cite{im06,wa09,ra17}
\begin{equation}
  A = \int_{A_p} dx \, dy \, \sqrt{1 + (\nabla h(x,y))^2}  \; ,
\label{aap}
\end{equation}
where $h(x,y)$ is the height of the membrane surface, i.e. the distance to 
the reference $(x, y)$ plane.
The difference between the expansion coefficients $\alpha$ and 
$\alpha_p$ at high temperature can be understood from the relation
between real and projected areas, $A$ and $A_p$, given by this equation.

In fact, the difference $A - A_p$ for a continuous membrane 
in a classical approach may be calculated by Fourier transformation 
of the r.h.s. of Eq.~(\ref{aap}).\cite{sa94,ch15,ra17}
This requires the introduction of a dispersion relation $\omega({\bf k})$
for out-of-plane modes (ZA band), where ${\bf k} = (k_x, k_y)$ are
2D wavevectors. The frequency dispersion in this acoustic (flexural)
band can be well approximated by the expression  
$\rho \, \omega^2 = \sigma k^2 + \kappa k^4$,
consistent with an atomic description of graphene\cite{ra16}
($k = |{\bf k}|$; $\rho$, surface mass density; 
$\sigma$, effective stress; $\kappa$, bending modulus).
Thus, one finds\cite{ra17}
\begin{equation}
 A = A_p  \left( 1 + \frac{k_{B} T}{4 \pi}
      \int_0^{k_{m}} d k \frac {k} {\sigma + \kappa k^2} \right)  \,  ,
\label{aap2}
\end{equation}
with the wavevector cut-off $k_{m} = (2 \pi /A_p)^{1/2}$.

For effective stress $\sigma > 0$, which is the case at finite 
temperatures, even for zero external stress ($P = 0$),\cite{ra16,ra17} 
the integral in Eq.~(\ref{aap2}) converges, yielding
\begin{equation}
 A = A_p \left[ 1 + \frac {k_{B} T} {8 \pi \kappa}
      \ln \left(1 + \frac{2 \pi \kappa} {\sigma A_p} \right) \right]  \:.
\label{aap3}
\end{equation}
Note that this expression has been derived in the classical limit, i.e.,
without taking into account atomic quantum delocalization. It is
expected, however, to be a good approximation to our quantum calculations
at relatively high temperature, $T \gtrsim \Theta_D$, with  
$\Theta_D \sim$ 1000 K the Debye temperature associated to 
out-of-plane vibrations in graphene.\cite{te09,po11}

It is not straightforward to write down an analytical expression for
$\alpha - \alpha_p$ from a temperature derivative of Eq.~(\ref{aap3}),
as one has to include changes in $\kappa$ and $\sigma$ through
$\partial \kappa / \partial T$ and $\partial \sigma / \partial T$.
One can instead obtain the temperature dependence of these parameters
from a fit to earlier results of classical simulations.\cite{ra16}
Thus, we obtain from Eq.~(\ref{aap3}) at $T$ = 1000~K a difference
$\alpha - \alpha_p = 8.5 \times 10^{-6}$ K$^{-1}$, close to the
high-temperature results obtained from our PIMD simulations
($\sim 10^{-5}$ K$^{-1}$).

Our low-temperature data for $A_p(T)$ and the trend $\alpha_p \to 0$ 
in the low-temperature limit are consistent with the results
obtained by Amorim {\em et al.},\cite{am14} from first-order perturbation 
theory and a one-loop self-consistent approximation.
These authors emphasized that the limits $N \to \infty$ and $T \to 0$ 
do commute, which agrees with the results of our simulations,
i.e., at low $T$ all system sizes yield the same results.
In general, the evaluation of low-temperature properties from PIMD
simulations becomes increasingly harder for both, larger $N$ and 
lower $T$. In the case of graphene, this is complicated by the fact 
that larger sizes may require lower temperatures to converge to the
ground-state properties, as shown for the area $A_p$ in Fig.~4.

\section{Specific heat}

\subsection{Harmonic approximation}

For comparison with the results of our PIMD simulations for the 
specific heat of graphene, we will discuss here a harmonic approximation
(HA) for the lattice vibrations. Even though this approximation will
turn out to be rather accurate at low temperatures, it is clear that
anharmonicity will show up as temperature is raised, and the results
of this approximation will progressively depart from those more 
realistic derived from the simulations. The HA assumes constant frequencies
for the graphene vibrations (those derived for the minimum-energy 
configuration), and does not take into account changes of the 
areas $A$ and $A_p$ with temperature. For solids, volume changes are usually
considered through quasi-harmonic approximations, which take into account
the thermal expansion and its corresponding changes in vibrational
frequencies (usually by means of Gr\"uneisen constants\cite{as76,mo05,ra12}). 
The same procedure is not directly applicable for graphene at any 
temperature, as due to the compression of $A_p$ the crystalline membrane 
becomes unstable in a quasi-harmonic approximation for $A_p < A_0$ 
with the appearance of imaginary frequencies when diagonalizing 
the dynamical matrix. 
This can be remedied at low temperatures for finite-size graphene layers, 
but the whole scheme becomes unstable at relatively high $T$, or yields 
unphysical results, as a continuous contraction of graphene at any 
temperature,\cite{mo05} in disagreement with results of both classical
and PIMD simulations.\cite{za09,mi15b,lo16,ga14,he16}

For a simulation cell including $N$ atoms, the specific heat
per atom, $c_v(T) = d E(T) / d T$, is given in the HA by
\begin{equation}
 c_v(T) = \frac {k_B}{N} \sum_{r,\bf k}
   \frac { \left[ \frac12 \beta \hbar \, \omega_r({\bf k}) \right]^2 }
   { \sinh^2 \left[ \frac12 \beta \hbar \, \omega_r({\bf k}) \right] } \, ,
\label{cvn}
\end{equation}
where $\beta = 1 / (k_B T)$, and 
the index $r$ ($r$ = 1, ..., 6) refers to the six phonon bands
of graphene (ZA, ZO, LA, TA, LO, and TO).\cite{mo05,ka11,wi04}
The sum in ${\bf k}$ is extended to wavevectors
${\bf k} = (k_x, k_y)$ in the hexagonal Brillouin zone,
with discrete ${\bf k}$ points spaced by $\Delta k_x = 2 \pi / L_x$ and
$\Delta k_y = 2 \pi / L_y$.\cite{ra16}
Eq.~(\ref{cvn}) has been used to calculate the specific heat presented below.
Increasing the system size $N$ causes the appearance of vibrational modes
with longer wavelength $\lambda$. In fact, one has for the phonons 
an effective cut-off
$\lambda_{max} \approx L$, with $L = (N A_p)^{1/2}$,
and the minimum wavevector is
$k_0 = 2 \pi / \lambda_{max}$, i.e., $k_0 \sim N^{-1/2}$.

The low-temperature behavior of the heat capacity vs $T$ can be
further analyzed by considering a continuous model for frequencies and
wavevectors, similarly to the well-known Debye model for 
solids.\cite{as76,ki66}
At low-temperatures, $c_v$ is dominated by the contribution of 
acoustic modes with small $k$. For graphene, these are TA and LA modes 
with $\omega_r \propto k$ and ZA modes with $\omega_r \propto k^2$
($\sigma$ is negligible at low $T$ and zero external stress). 

The low-$T$ contribution of a phonon branch with dispersion 
relation $\omega_r \propto k^n$ can be approximated as
\begin{equation}
 c_v^r(T) \approx   k_B  \int_{k_0}^{k_m}
    \frac { \left[ \frac12 \beta \hbar \, \omega_r(k) \right]^2 }
      { \sinh^2 \left[ \frac12 \beta \hbar \, \omega_r(k) \right] } \,
             \rho(k) \, d k  \, ,
\label{cvr}
\end{equation}
where $k_m$ is the maximum wavenumber
$k_m = (2 \pi / A_0)^{1/2}$  and
$\rho(k) = A_0 k / 2 \pi$ for 2D systems.
From the dispersion relation $\omega_r(k)$, we have
a vibrational density of states 
$\bar{\rho}_r(\omega) \sim \omega^{\frac2n - 1}$
Introducing the large-size limit $\omega_0 \to 0$ and
putting $x = \frac12 \beta \hbar \, \omega$, one finds
\begin{equation}
c_v^r \sim   k_B  \frac{K}{(\beta \hbar)^{\frac2n}}
     \int_0^{x_m}  \frac { x^{\frac2n + 1} } { \sinh^2 x } \, d x   \, ,
\label{cvr3}
\end{equation}
$K$ being a constant.
Then, for low temperatures, $k_B T \ll \hbar \, \omega_m$, 
one has $c_v^r \sim T^{2/n}$.
In general, for $d$-dimensional systems one finds
an exponent $d/n$.\cite{ho01,po02}
Thus, for the ZA phonon branch in graphene ($n = 2$), one expects
a linear dependence of $c_v^r$ on $T$, whereas
for the contribution of LA and TA branches ($n = 1$), one has
at low temperature $c_v^r \sim T^2$.
Putting all constants in the integrals, we find for the ZA branch
\begin{equation}
   c_v^{ZA} = \frac{\pi}{12} \frac{k_B^2}{\hbar}  
             \sqrt{\frac{\rho}{\kappa}}  A_0 \, T  \, ,
\label{cvt}
\end{equation} 
and for acoustic LA and TA modes:
\begin{equation}
   c_v^{ac} = \frac{3 \zeta(3)}{\pi}  \frac{k_B^3}{\hbar^2}
               \frac{A_0}{v^2} \, T^2  \, ,
\label{cvt2}
\end{equation}
$v$ being the sound velocity in the corresponding phonon branch,
and $\zeta$ is the Riemann zeta function.

For in-plane vibrations of graphene, the acoustic branches can be described 
by the linear dispersion $\omega_i = v_i k$, with sound speed
$v_1$ = 21.5 km/s for LA and $v_2$ = 14.0 km/s for TA modes. These values
for $v_1$ and $v_2$ were derived from the elastic properties of
graphene obtained by using the LCBOPII potential,\cite{ra17} and are close 
to those given by Karssemeijer and Fasolino,\cite{ka11} as well as to those
derived from {\em ab-initio} calculations for graphene\cite{mo05} and
experimental data for graphite.\cite{wi04}
For the bending constant $\kappa$ describing the ZA phonon band,
we take $\kappa$ = 1.49 eV.\cite{ra16}

\subsection{Elastic energy}

Apart from the pure vibrational energy associated to the phonons in
graphene, one has even in the presence of an externally applied stress,
an elastic energy due to changes in the area $A$ of the crystalline
membrane (thermal expansion). Thus, the internal energy $E(T)$ 
at temperature $T$ can be written as\cite{he16}
\begin{equation}
    E(T) =  E_0 + E_{\rm el}(A) + E_{\rm vib}(A,T)   \, ,
\label{et}
\end{equation}
where $E_{\rm el}(A)$ is the elastic energy corresponding to an
area $A$, and $E_{\rm vib}(A,T)$ is the vibrational energy of the system.
Our PIMD simulations directly give $E(T)$, which can be split
into an elastic and a vibrational part.

\begin{figure}
\vspace{-1.0cm}
\includegraphics[width=8.5cm]{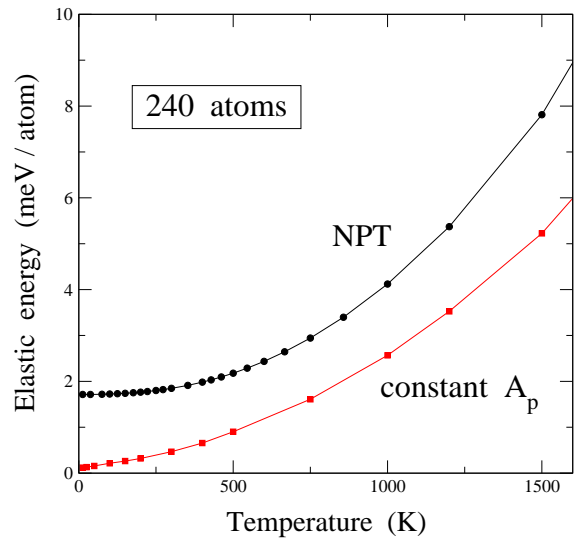}
\vspace{-0.5cm}
\caption{Elastic energy of graphene, $E_{\rm el}$, as derived from
the real area obtained in PIMD simulations in the $NPT$ ensemble for
$P = 0$ (circles) and in the constant-$A_p$ ensemble with the in-plane
area corresponding to the minimum energy, $A_p = A_0$ (squares).
Lines are guides to the eye.
}
\label{f7}
\end{figure}

The elastic energy $E_{\rm el}$ corresponding to an area $A$ is defined 
here as the increase in energy of a strictly planar graphene layer with 
respect to the minimum energy $E_0$.
We have calculated $E_{\rm el}(A)$ for a supercell including 960 atoms,
expanding it isotropically and keeping it flat.
As indicated above for the area $A$, finite-size effects on the
elastic energy are very small, and in practice negligible for our
current purposes.
The elastic energy $E_{\rm el}(A)$ increases with $A$, and for small
lattice expansion it can be approximated as
$E_{\rm el}(A) \approx K (A - A_0)^2$, with $K$ = 2.41 eV/\AA$^2$.
The elastic energy turns out to be much smaller than the vibrational energy
in the cases considered here, but it can be nonnegligible for
the actual heat capacity of graphene, as the area $A$ changes with
temperature.

In Fig.~7 we present the elastic energy $E_{\rm el}$ as a function of
temperature for our $NPT$ simulations at $P = 0$.
For comparison we also show results derived from constant-$A_p$ simulations 
with the minimum-energy area $A_0$.
$E_{\rm el}$ is found to increase with $T$ in both cases, but there are
some differences between them.  In the isothermal-isobaric 
simulations (circles) we find an appreciable elastic energy in the 
zero-temperature limit, basically due to zero-point lattice expansion 
(see Sec.~IV). In the constant-$A_p$ simulations (squares) such 
an expansion is not allowed, and 
$E_{\rm el}$ is nearly zero; a small positive value of
$E_{\rm el}$ is obtained for $T \to 0$, caused by a slight increase in the
area $A$ due to out-of-plane zero-point vibrations.
The difference between elastic energy in both kinds of simulations
decreases from $T$ = 0 until about 1000 K, and then it increases at
higher temperatures. This is related to the dependence of the in-plane area
in $NPT$ simulations upon temperature, which approaches the area
$A_0$ up to 1000 K, and departs from it at higher $T$ (see Fig.~4).

\begin{figure}
\vspace{-1.0cm}
\includegraphics[width=8.5cm]{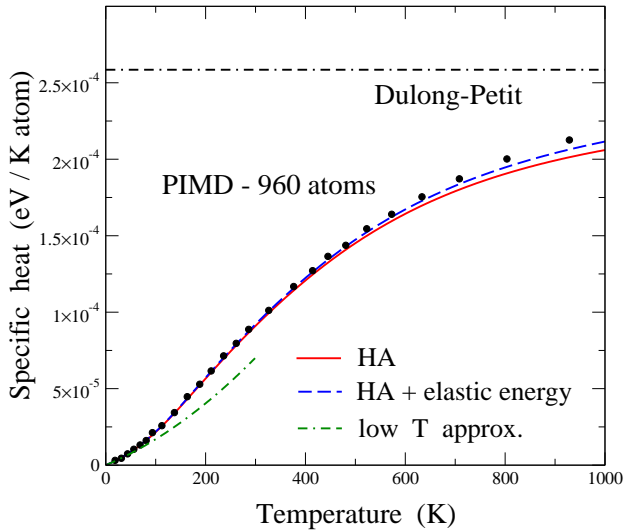}
\vspace{-0.5cm}
\caption{Specific heat of graphene as a function of temperature.
Symbols represent results for $c_p$ derived from PIMD simulations
for $N$ = 960 atoms.
Error bars of the data points are less than the symbol size.
The solid line is $c_v$ obtained from the six phonon
bands corresponding to the LCBOPII potential in a harmonic approximation.
The dashed line includes the contribution $d E_{\rm el} / d T$ of the
elastic energy.
The horizontal dashed-dotted line represents the harmonic classical limit
(Dulong-Petit law).
}
\label{f8}
\end{figure}

\subsection{Comparison with PIMD simulations}

We now turn to the results of our simulations.
In Fig.~8 we show the specific heat of graphene as a function of temperature.
Solid circles represent results for $c_p$, derived from PIMD simulations
for $N$ = 960 atoms. They were obtained from a
numerical derivative of the internal energy $E(T)$. Results corresponding
to $N$ = 240 are indistinguishable from those plotted in Fig.~8.
The solid line was calculated with the HA using Eq.~(\ref{cvn}),
with the frequencies $\omega_r({\bf k})$ ($r$ = 1, ..., 6) obtained from 
diagonalization of the dynamical matrix corresponding to the 
LCBOPII potential.\cite{ka11}
Results of the simulations follow closely the HA up to about 400~K, and 
they become progressively higher than the solid line for higher temperatures, 
at which anharmonic effects are expected to increase.

As indicated above, a part of the internal energy at a given 
temperature corresponds to the elastic energy $E_{\rm el}$, 
i.e. to the cost of increasing the area $A$ of graphene. 
We have calculated the contribution of this energy to the
specific heat as $d E_{\rm el} / d T$, using the data obtained from 
PIMD simulations in the $NPT$ ensemble, shown in Fig.~7.
To assess the importance of this contribution to the whole
specific heat, we have added it to the result of the HA
(solid line in Fig.~8), and have displayed the sum as a dashed line.
We find an observable increase in the specific heat respect the pure HA, 
especially visible for $T > 500$ K, such that it incorporates
part of the anharmonicity of the system, yielding a result closer
to the specific heat $c_p$ derived from the simulations.
For comparison, the classical Dulong-Petit specific heat is shown as
a dashed-dotted line ($c_v^{cl} = 3 k_B$).
At $T$ = 1000 K the quantum results are still appreciably lower than the
classical limit. 

We have also calculated the specific heat $c_v$ from constant-$A_p$ 
simulations.  For each temperature, we take the equilibrium area $A_p$ 
obtained in the $NPT$ simulations, and calculate $c_v$ as 
$\Delta E / \Delta T$ for that value of $A_p$ from increments 
$\Delta T$ (both positive and negative). 
Note that this is not the same as taking a temperature derivative of
the energy curve shown in Fig.~1 for the constant-$A_p$ ensemble, since in
this case the simulations were carried out with minimum-energy area $A_0$.
It is expected that $c_v \leq c_p$ at any temperature, but the difference 
between them turns out to be smaller than the statistical error bar 
of our results, so that they appear as indistinguishable from the 
direct results of our PIMD simulations (see below).

\begin{figure}
\vspace{-1.0cm}
\includegraphics[width=8.5cm]{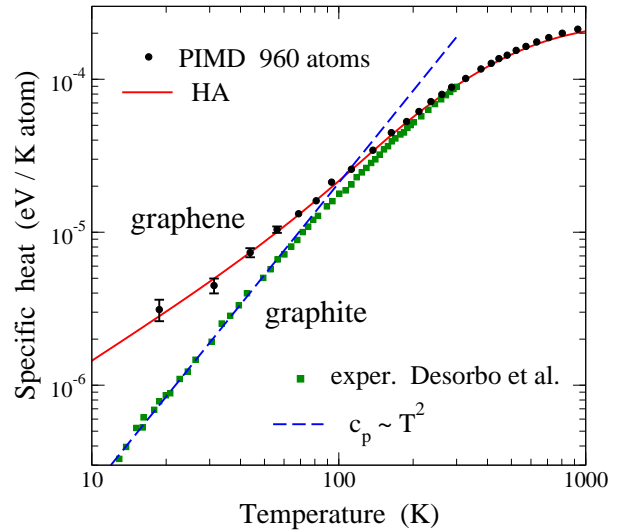}
\vspace{-0.5cm}
\caption{Specific heat of graphene as a function of temperature.
Solid circles represent results for $c_p$ derived from PIMD simulations
for $N$ = 960 atoms. The solid line was obtained from the six phonon
bands corresponding to the LCBOPII potential in a harmonic approximation.
For comparison, experimental data for graphite obtained by
Desorbo and Tyler\cite{de53} are shown as squares.
The dashed line displays a dependence $c_p \sim T^2$.
}
\label{f9}
\end{figure}

An analytical expression for $c_v(T)$ at low temperature can be derived
in the HA from the contribution $c_v^r$ of modes ZA, LA, and TA,
as given by Eqs.~(\ref{cvt}) and (\ref{cvt2}).
The sum of contributions of these three phonon bands is plotted in
Fig.~8 as a dashed-dotted line. It follows the simulation data and the
whole HA up to $T \sim$ 50 K, and at high temperatures it becomes lower.

To display more clearly the low-temperature region, we present
in Fig.~9 the specific heat vs temperature in a semilogarithmic plot.
Solid circles are results for $c_p$ derived from PIMD simulations for 
$N$ = 960.
The solid line indicates the harmonic approximation for $c_v$ obtained 
from Eq.~(\ref{cvn}) for the same cell size.
At low temperature, it is clear the linear dependence of the
specific heat on $T$ (slope unity in the logarithmic plot).
In fact we find at low temperature $c_p \approx C T$ with
$C = 1.4 \times 10^{-7}$ eV K$^{-2}$
(Note that $C = 2 a$, with $a$ the coefficient of the quadratic term
in the fit of the energy shown in Fig.~3).
For comparison we also present in Fig.~9 experimental data for $c_p$ of
graphite, obtained by Desorbo and Tyler.\cite{de53}
The temperature dependence of the heat capacity of graphite has 
been studied in detail along the years.\cite{ko51,kr53,kl53,ni72}
In this case, $c_p$ rises as $T^3$ for $T <$ 10~K (a region
not reached here and not shown in Fig.~9). At temperatures  between
10 and 100~K, $c_p$ increases as $T^2$.
The main difference with graphene is that the dominant contribution to
$c_p$ in this temperature region comes from phonons with a linear 
dispersion relation ($\omega \sim k$) for small $k$.
At room temperature the specific heat of graphite amounts to  
$8.90 \times 10^{-5}$ eV / K atom, i.e. 8.59 J / K mol,
somewhat smaller than our results of PIMD simulations for
graphene: $c_p = 9.4(\pm 0.1) \times 10^{-5}$ eV / K atom.

The difference between $c_p$ and $c_v$ has been obtained from
the thermodynamic relation
\begin{equation}
  c_p - c_v = T  \alpha_p^2  B_p  A_p  \,
\label{cpcv}
\end{equation}
where $B_p$ is the in-plane isothermal bulk modulus, i.e.
$B_p = - A_p \, (\partial P / \partial A_p)_T$.
Eq.~(\ref{cpcv}) is similar to the relation between $c_p$ and $c_v$
for 3D systems.\cite{ca60,la80}
Note that the variables appearing on the r.h.s. of Eq.~(\ref{cpcv})
refer to in-plane properties, as the pressure appearing in our $NPT$ 
ensemble is the conjugate variable of the in-plane area $A_p$.
$B_p$ has been calculated by using the fluctuation formula:\cite{la80,ra17}
\begin{equation}
   B_p = \frac{k_B T A_p}{N (\Delta A_p)^2} 
\label{bp}
\end{equation}
with $(\Delta A_p)^2$ the mean-square fluctuations of the in-plane 
area $A_p$ obtained in the simulations.
This expression is more convenient for our purposes than obtaining
$(\partial A_p / \partial P)_T$, as this derivative requires additional 
simulations at nonzero stresses. In any case, we have checked
at some selected temperatures that both procedures yield the same
results for $B_p$ (taking into account the error bars).

\begin{figure}
\vspace{-1.0cm}
\includegraphics[width=8.5cm]{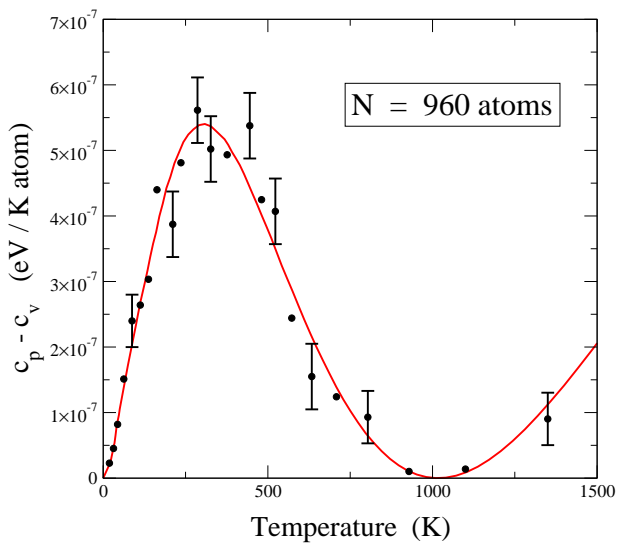}
\vspace{-0.5cm}
\caption{Difference $c_p - c_v$ vs temperature, as obtained from the
thermodynamic expression $c_p - c_v = T \alpha_p^2 B_p A_p$.
Symbols were derived from the values of $\alpha_p$, $B_p$, and
$A_p$ directly derived from PIMD simulations, whereas the solid line was
calculated from polynomial fits of these quantities as a function of
temperature.
}
\label{f10}
\end{figure}

Thus, we have obtained the difference $c_p - c_v$ introducing into
Eq.~(\ref{cpcv}) the values of $\alpha_p$, $B_p$, and  $A_p$ derived from
PIMD simulations. The results are shown in Fig.~10 as a function of 
temperature (solid circles).
The solid line in this figure was obtained by using Eq.~(\ref{cpcv})
and polynomial fits for the temperature dependence of the factors
in the r.h.s. of this equation. 
In the zero-temperature limit the difference $c_p - c_v$ converges to zero, 
as should happen because both specific heats vanish as $T \to 0$, and this 
difference increases for rising temperature until $T \sim$ 300 K, where it 
reaches a maximum
of  $\approx 5.5 \times 10^{-7}$ eV/(K atom). At higher $T$,
$c_p - c_v$ decreases and reaches zero at the temperature at which
$\alpha_p$ vanishes ($\sim$ 1000~K).
Note that the vanishing of the difference $c_p - c_v$ at a finite 
temperature (here $T \sim$ 300~K) obtained for graphene is similar
to that occurring for some 3D materials with negative thermal expansion 
at low $T$ (e.g., crystalline silicon),\cite{sh72,ev99} according to
the thermodynamic equation $c_p - c_v = T \alpha^2 B V$.

We emphasize here that the calculation of the low-temperature specific
heat of materials using path-integral simulations is not in general an 
easy task. In fact, even the reproduction of the Debye law $c_p \sim T^3$
for 3D solids has been a challenge for PIMD of solids, due to the effective
low-frequency cut-off associated to finite simulation cells.\cite{no96,ra06}
The reliability of this kind of calculations for 2D materials such as
graphene is mainly due to two reasons.
First, the length of the cell sides scales as $L \sim N^{1/d}$, so that
the minimum wavevector accessible in the simulation is $k_0 \sim N^{-1/d}$. 
Then, for a simulation cell including $N$ atoms, 
$k_0$ is smaller for 2D than for 3D materials,
which means that the low-frequency region is better described in the 
former case, and consequently also the low-temperature regime. 
Second, and even more important, is the fact that the internal energy
for graphene rises as $T^2$ (or $c_p \sim T$), which is a fast increase,
much easily detectable than the typical expectancy ($E \sim T^4$) for the 
phonon contribution in 3D materials ($c_p \sim T^3$).

We note that the electronic contribution to the specific heat of graphene
has not been taken into account for the temperatures considered here, 
as it is much less than the phonon contribution. 
The former has been estimated in several works, and
turns out to be between three and four orders of magnitude smaller than 
the latter in pure graphene.\cite{be96,ni03,fo13}

A thermodynamic parameter related to the thermal expansion and specific
heat is the dimensionless Gr\"uneisen parameter $\gamma$,\cite{as76} 
which for our in-plane variables in graphene can be written as
\begin{equation}
       \gamma = \frac{B_p \alpha_p A_p}{c_v}  \, .
\end{equation}
From the results of our PIMD simulations for $N$ = 960 atoms, we
find $\gamma = -2.3$ at 300~K, and at 1000~K, $\gamma \approx 0$
within the precision of our numerical results.
At room temperature $\gamma$ and $\alpha_p$ turn out to be negative
mainly as a consequence of the negative sign of the mode-dependent
Gr\"unesien parameter $\gamma_{\rm ZA}$ for the out-of-plane ZA
vibrations.\cite{ba11,mo05} At temperatures in the order of 1000~K
this trend is compensated for by the positive sign of the Gr\"unesien 
parameters of in-plane modes, which eventually causes that 
the overall $\gamma$ and $\alpha_p$ become positive for $T >$ 1000~K.

\section{Summary}

We have presented results of PIMD simulations of graphene monolayers 
in the isothermal-isobaric ensemble at several temperatures and
zero external stress.
Consideration of quantum dynamics of the atomic nuclei has allowed
us to realistically describe structural and thermodynamic properties
of graphene at finite temperatures. Such a quantum description is
crucial to study thermal properties at temperatures in the order of
and below room temperature.

The LCBOPII potential model describes fairly well the vibrational 
frequencies of graphene. We have shown here that 
quantum effects associated to vibrational motion are also 
described in a reliable manner by PIMD simulations using this potential.

We have discussed the fact that the so-called thermal contraction of 
graphene presented in the literature is in fact a decrease in the in-plane 
(projected) area $A_p$ due to out-of-plane vibrations, and not to 
a reduction in the real area $A$ of the graphene sheet.
The difference $A - A_p$ grows as temperature is raised,
because of the larger amplitude of those vibrations.
The in-plane thermal expansion $\alpha_p$ is found to be negative at low 
temperature, and becomes positive for $T \gtrsim$ 1000~K. 
However, the thermal expansion $\alpha$ of the real area turns out to be 
positive at all finite temperatures.

Anharmonicity of the vibrational modes is appreciable and should be
taken into account in any finite-temperature calculation of the properties
of graphene. This manifests itself clearly in the temperature dependence
of the in-plane and real areas shown in Fig.~4.
However, other thermal properties of graphene are well described
by the HA once the frequencies of the vibrational modes
are known for the classical equilibrium geometry at $T = 0$.
A calculation of the specific heat $c_p$ from results of PIMD simulations 
indicates that anharmonicity shows up progressively at temperatures
$T \gtrsim$ 400~K.
In particular, a contribution to the heat capacity, not included in the
HA, comes from the elastic energy associated to 
the expansion of the actual graphene sheet at finite $T$ ($\alpha > 0$).
At the lowest temperatures studied here ($T >$ 10~K) we find a linear
dependence of the specific heat of graphene $c_p = C T$, with 
$C = 1.4 \times 10^{-7}$ eV K$^{-2}$.

PIMD simulations similar to those presented here may help
to understand thermal properties of graphane (a hydrogen monolayer on
graphene), as well as the dynamics of free-standing graphene
multilayers.   \\  \\

The authors acknowledge the help of J. H. Los in the implementation 
of the LCBOPII potential.
This work was supported by Direcci\'on General de Investigaci\'on,
MINECO (Spain) through Grants FIS2012-31713 and FIS2015-64222-C2.


\end{document}